\documentstyle[preprint,revtex,eqsecnum]{aps}
\begin{document}
\preprint{PITT 93-01}
\draft
\begin{title}
{\bf Quantum Spinodal Decomposition}
\end{title}
\author{{\bf Daniel Boyanovsky}}
\begin{instit}
Department of Physics and Astronomy\\ University of
Pittsburgh \\
Pittsburgh, P. A. 15260, U.S.A.
\end{instit}

\begin{abstract}
We study the process of spinodal decomposition in a scalar
quantum
field theory that is quenched from an equilibrium disordered
initial
state at $T_i > T_f$ to a final state at $T_f \approx 0$.
The process
of formation and growth of correlated domains is studied in
a Hartree
approximation. We find an approximate scaling law for the
size of the
domains $\xi_D(t) \approx \sqrt{t \xi_0}$ at long times for
weakly
coupled theories, with $\xi_0$ the zero temperature
correlation length.
\end{abstract}
\pacs{11.10.-z;11.9.+t;64.90.+b}
\narrowtext
The process of phase separation through spinodal
decomposition is well
understood  within the context of classical non-equilibrium
statistical
mechanics\cite{langer1,guntonmiguel,nigel}.
When the quench is at criticality, it is primarily
associated with the onset and growth of unstable long-
wavelength modes
(we will not discuss nucleation here).

Spinodal decomposition is conjectured to have taken place
during phase
transitions in the early universe typically described by
scalar quantum
field theories\cite{brandenberger,kolb,linde}.
 Thus it becomes an important issue to
understand and
describe the mechanism of phase separation in quantum field
theory with
the definite motivation provided by cosmological phase
transitions. An attempt to study some of these issues was reported by
 Mazenko and collaborators\cite{mazenko}.
Understanding the process of phase separation in quantum
many-body systems
may also prove relevant within the context of mesoscopic
systems and quantum phase transitions in condensed matter. In this
article, we introduce the techniques of non-equilibrium
quantum statistical
mechanics to study the process of phase separation in a
typical scalar
field theory.

{}From the outset we recognize fundamental differences between
classical
theories of spinodal decomposition, and the description of
phase
separation in a quantum system. In the former, the equations
of motion are
purely dissipative, whereas in the latter, the Heisenberg
operator
equations of motion are second order and thus time reversal
invariant.
Furthermore, thermal fluctuations are incorporated in the
classical
description via a Langevin noise term, typically
uncorrelated and
satisfying the fluctuation-dissipation relation, whereas in
the quantum
description, both thermal and quantum fluctuations are
present in the
initial {\it density matrix} that describes the initial
ensemble.

We model the physical situation of a ``critical quench'' via
a time
dependent Hamiltonian
\widetext
\begin{eqnarray}
  H(t) & = & \int_{\Omega} d^3x \left\{
\frac{1}{2}\Pi^2(x)+\frac{1}{2}(\vec{\nabla}\Phi(x))^2+\frac
{1}{2}m^2(t)
\Phi^2(x)+\frac{\lambda}{4!}\Phi^4(x) \right \}
\label{timedepham} \\
m^2(t) & = & m^2_i \Theta(-t) - m^2_f \Theta(t)
\label{massoft} \\
m^2_i  & = & \mu^2[\frac{T^2_i}{T^2_c}-1] \; \; ; \; \;
m^2_f =  \mu^2[1-\frac{T^2_f}{T^2_c}] \label{massfinal}
\end{eqnarray}
\narrowtext
with $ \mu^2 > 0 \; ; \; T_i > T_c \; ; \; T_f \approx 0$.
The initial state of the system (at $t<0$) is assumed to be
described by
an equilibrium density matrix at the initial temperature
$T_i$.
\widetext
\begin{eqnarray}
\hat{\rho}_i  & = & e^{-\beta_i H_i} \label{initialdesmat}\\
          H_i & = & H(t<0) \label{initialham}
\end{eqnarray}
\narrowtext

In the Schroedinger picture, the density matrix evolves in
time as
\widetext
\begin{equation}
\hat{\rho(t)} = U(t)\hat{\rho}_iU^{-1}(t) \label{timedesmat}
\end{equation}
\narrowtext
with $U(t)$ the time evolution operator. In the present case, the
order parameter ($Tr \rho(t)\int_{\Omega}d^3x \Phi(\vec{x})$) obeys a
Heisenberg equation of motion and is {\it not conserved}.
The expectation value of operators may be computed by
introducing the
generating functional
\widetext
\begin{equation}
Z[J^+, J^-]= Tr U(T-i\beta_i,T)
U(T,T';J^-)U(T',T;J^+)\label{generatingfunctional}
\end{equation}
\narrowtext
with $T \rightarrow -\infty ; \; ; T' \rightarrow \infty$.
It is found
\widetext
\begin{eqnarray}
Z[J^+,J^-] & = & \exp{\left\{i\int_{T}^{T'}dt\left[
{\cal{L}}_{int}(-i\delta/\delta J^+)-
{\cal{L}}_{int}(i\delta/\delta
J^-)\right] \right \}} \times \nonumber \\
           &   & \exp{\left\{\frac{i}{2}\int_T^{T'}
dt_1\int_T^{T'} dt_2 J_a(t_1)J_b(t_2)G_{ab} (t_1,t_2)
\right\}}
 \label{generatingfunction}
\end{eqnarray}
\narrowtext
with $G_{ab}$ the Green's function on a contour\cite{calzetta}.
The quantities of interest are
\widetext
\begin{eqnarray}
S(\vec{r};t) & = &
\langle\Phi(\vec{r},t)\Phi(\vec{0},t)\rangle
\label{equaltimecorr} \\
S(\vec{r};t) & = & \int \frac{d^3 k}{(2\pi)^3}e^{i\vec{k}
\cdot \vec{r}}
\langle\Phi_{\vec{k}}(t)\Phi_{-\vec{k}}(t)\rangle \label{strufack}
 \label{strucfac}
\end{eqnarray}
\narrowtext
The equal time correlation function at zeroth order (``tree
level'')
 is thus found to be
\widetext
\begin{equation}
\langle\Phi_{\vec{k}}(t)\Phi_{-\vec{k}}(t)\rangle=
\frac{1}{2 \omega_<(k)}{\cal{U}}^+_k(t){\cal{U}}^-
_k(t)\coth\left[\beta_i
\omega_<(k)/2\right]
\end{equation}
\narrowtext
with the mode functions obeying
\widetext
\begin{eqnarray}
\left[\frac{d^2}{dt^2} + \vec{k}^2 +
m^2(t)\right]{\cal{U}}_k^{\pm} & = & 0 \label{homogeneous}
 \end{eqnarray}
\narrowtext
with $m^2(t)$ given by (\ref{massoft}).

The boundary conditions on the homogeneous solutions are
\widetext
\begin{equation}
{\cal{U}}_k^{\pm}(t<0) =  e^{\mp i
\omega_{<}(k)t} \; \; ; \; \; \omega_{<}(k) =
\left[\vec{k}^2+m^2_i\right]^{\frac{1}{2}}\label{boundaryconditions}
\end{equation}
\narrowtext
corresponding to positive frequency (particles) and negative
frequency (antiparticles) (${\cal{U}}_k^{+}(t);
{\cal{U}}_k^{-}(t)$, respectively).

The ``free field'' mode functions are easily found. For
$t>0$, they
consist of stable modes ($\vec{k}^2 > m_f^2$) and unstable
modes
($\vec{k}^2 < m_f^2$). These unstable modes are responsible
for the
growth of correlations. The zeroth order equal time
correlation function
becomes
\widetext
\begin{equation}
\langle\Phi_{\vec{k}}(t)\Phi_{-\vec{k}}(t)\rangle  =
 \frac{1}{2 \omega_{<}(k)}  \coth[\beta_i\omega_{<}(k)/2]
\end{equation}
\narrowtext
for $t< 0$, and

\widetext
\begin{eqnarray}
\langle\Phi_{\vec{k}}(t)\Phi_{-\vec{k}}(t)\rangle
 & = & \frac{1}{2 \omega_{<}(k)} \{
\left[1+2A_kB_k\left[ \cosh(2W(k)t)-1 \right]\right]
 \Theta(m^2_f-\vec{k}^2)
\nonumber  \\
 & + &
\left[1+2a_kb_k\left[ \cos(2 \omega_{>}(k)t)-1
\right]\right]
 \Theta(\vec{k}^2-m^2_f)\}
\coth[\beta_i\omega_{<}(k)/2]
\end{eqnarray}
\narrowtext
with $\omega_{>}(k)= \sqrt{\vec{k}^2-m_f^2}$ and $W(k)=
\sqrt{m_f^2-\vec{k}^2}$, for $t > 0$.

The first term, the contribution of the unstable modes,
reflects the growth of correlations because of the
instabilities and
will be the dominant term at long times in this approximation.

It is convenient to introduce the dimensionless quantities
\widetext
\begin{equation}
\kappa = \frac{k}{m_f}  \; \; ; \; \;  L^2 =
\frac{m_i^2}{m_f^2}=
\frac{\left[T^2_i-T^2_c\right]}{\left[T^2_c-T^2_f\right]}
\; \; ; \; \; \tau = m_f t  \; \; ; \; \;  \vec{x} = m_f
\vec{r}
\label{dimensionless}
\end{equation}
\narrowtext
and the critical temperature
 $T_c^2= 24 \mu^2 / \lambda$\cite{brandenberger,kolb,linde}, in
terms of
which the  ``tree level'' subtracted structure factor
$S^{(0)}(k,t)-
S^{(0)}(k,0)=(1/m_f){\cal{S}}^{(0)}(\kappa,\tau)$ becomes
\widetext
\begin{eqnarray}
{\cal{S}}^{(0)}(\kappa,\tau)   & = &
\left( \frac{24}{\lambda [1-\frac{T^2_f}{T^2_c}]}
\right)^{\frac{1}{2}}
\left(\frac{T_i}{T_c} \right)
\frac{1}{2\omega^2_{\kappa}} \left( 1+
\frac{\omega^2_{\kappa}}{W^2_{\kappa}}
\right) \left[ \cosh(2W_{\kappa}\tau)-1 \right]
\label{strucuns} \\
             \omega^2_{\kappa} & = & \kappa^2+L^2
\; \; ; \; \;   W_{\kappa}  =  1-\kappa^2
\label{bigomegak}
\end{eqnarray}
\narrowtext
To obtain a better idea of the growth of correlations, it is
convenient
to introduce the scaled correlation function
\widetext
\begin{equation}
{\cal{D}}(x,\tau) = \frac{\lambda}{6m^2_f}\int^{m_f}_0
\frac{k^2 dk}{2\pi^2}\frac{\sin(kr)}{(kr)}[S(k,t)-S(k,0)]
\label{integral}
\end{equation}
\narrowtext
The reason for this is that the minimum of the tree level
potential occurs
at $\lambda \Phi^2 /6 m^2_f =1$, and the inflexion
(spinodal) point,
at $\lambda \Phi^2 /2 m^2_f =1$, so that ${\cal{D}}(0,\tau)$
measures the
excursion of the fluctuations to the classical spinodal and
beyond as the correlations grow in time.

At large $\tau$ (time), $\kappa^2 {\cal{S}}(\kappa,\tau)$
has a
sharp peak at $\kappa_s= 1/ \sqrt{\tau}$ with amplitude
$\exp[2\tau]/\tau$ (see figure 1). We find for $x < \tau$
and $T_f \approx 0$
\widetext
\begin{eqnarray}
{\cal{D}}(x,\tau) & \approx & {\cal{D}}(0,\tau)
\exp[-\frac{x^2}{8\tau}]
\frac{\sin(x/ \sqrt{\tau})}{(x/ \sqrt{\tau})}
\label{strucfacxtau} \\
{\cal{D}}(0,\tau) & \approx & \left(\frac{\lambda}{12
\pi^3}\right)^
{\frac{1}{2}}\left(\frac{(\frac{T_i}{2 T_c})^3}{[
\frac{T^2_i}{T^2_c}-
1]}\right)\frac{\exp[2\tau]}{\tau^{\frac{3}{2}}}
\label{strucfactau}
\end{eqnarray}
\narrowtext

 Restoring dimensions, and recalling that the zero
temperature correlation
 length is $\xi(0) = 1/\sqrt{2} \mu$,
 we find that for $T_f \approx 0$ the amplitude of the
fluctuation inside a
 ``domain'' $\langle \Phi^2(t)\rangle$, and the ``size'' of
a  domain  $\xi_D(t)$ grow as
\widetext
\begin{equation}
 \langle \Phi^2(t)\rangle  \approx   \frac{\exp[\sqrt{2}t/
\xi(0)]}
 {(\sqrt{2}t/ \xi(0))^{\frac{3}{2}}} \; \; ; \; \;
 \xi_D(t)                  \approx
(8\sqrt{2})^{\frac{1}{2}}
 \sqrt{t\xi(0)} \label{domainsize}
 \end{equation}
\narrowtext

The presence of the instabilities precludes a well-defined
perturbative
expansion. Consider a one loop
contribution to the
equal time correlation function. The ``external legs''
obtain a
contribution from the unstable modes, but in the loop
integral, the
integration over the momenta also includes a contribution
from the unstable modes.
It is clear
that eventually the one-loop correction dominates and
perturbation theory
breaks down, even for the case of very weak couplings.
This feature will persist to all orders in a perturbative expansion.
The {\it dynamics} of the phase transition {\it cannot be studied in
perturbation theory}.

Our non-perturbative approach is based on a Hartree
approximation, which
is similar to the early approach of  Langer\cite{langer2} for classical
spinodal decomposition. It is implemented by the replacement

\[m^2(t) \rightarrow
m^2(t)+\frac{\lambda}{2}\langle\Phi^2(t)\rangle \]
(where we used spatial translational invariance).

 This
leads to the self consistent set of equations
\widetext
\begin{equation}
\left[\frac{d^2}{dt^2}+\vec{k}^2+m^2(t)+\frac{\lambda}{2}\langle
\Phi^2(t)\rangle\right]
{\cal{U}}^{\pm}_k=0 \label{hartree}
\end{equation}
\narrowtext
\widetext
\begin{equation}
\langle\Phi^2(t)\rangle  =
\int \frac{d^3k}{(2\pi)^3} \frac{1}{2\omega_{<}(k)}
{\cal{U}}^+_k(t)
{\cal{U}}^-_k(t) \coth[\beta_i\omega_{<}(k)/2] \label{fi2}
\end{equation}
\narrowtext

The composite operator $\langle \Phi^2(\vec{r},t) \rangle$
needs one subtraction and multiplicative renormalization.
The subtraction
is absorbed in a renormalization of the bare mass, and the
multiplicative
renormalization into a renormalization of the coupling
constant.
The Hartree approximation provides a self-consistent non-
perturbative
scheme that sums an infinite series of Feynman diagrams\cite{chang}. For
$t<0$
there is a self-consistent solution  given by
\widetext
\begin{eqnarray}
\langle\Phi^2(t)\rangle       & = &  \langle\Phi^2(0)\rangle
\; \; ; \; \; {\cal{U}}^{\pm}_k(t)    =   \exp[\mp i
\omega_{<}(k)t] \\
\omega^2_{<}(k)               & = &
\vec{k}^2+m^2_i+\frac{\lambda}{2}+
\langle\Phi^2(0)\rangle = \vec{k}^2+m^2_{i,R} \nonumber \\
\end{eqnarray}
\narrowtext
and $m_{i,R}^2 = \mu_R^2[(T_i^2/T_c^2)-1]$.
For $t>0$ we subtract the composite operator at $t=0$
absorbing the subtraction
into a renormalization of $m^2_f$ which we now parametrize
as $m^2_{f,R}=
\mu^2_R[1-(T^2_f/T^2_c)]$. This
choice of
parametrization only represents a choice of the bare
parameters. The logarithmic
multiplicative
divergence of the composite operator will be absorbed in a
coupling constant
renormalization consistent with the Hartree
approximation\cite{chang}.
However, for the purpose of understanding the dynamics of
growth of
instabilities associated with the long-wavelength
fluctuations,
we will not need to specify this procedure. After
this renormalization, the Hartree equations read
\widetext
\begin{equation}
[\langle\Phi^2(t)\rangle-\langle\Phi^2(0)\rangle]  =
\int \frac{d^3k}{(2\pi)^3} \frac{1}{2\omega_{<}(k)}
[{\cal{U}}^+_k(t)
{\cal{U}}^-_k(t)-1] \coth[\beta_i\omega_{<}(k)/2]
\label{subfi2}
\end{equation}
\narrowtext
\widetext
\begin{equation}
\left[\frac{d^2}{dt^2}+\vec{k}^2+m^2_R(t)+\frac{\lambda_R}{2
}
\left(\langle\Phi^2(t)\rangle-\langle\Phi^2(0)\rangle\right)
\right]
{\cal{U}}^{\pm}_k(t)=0 \label{subhartree}
\end{equation}
\narrowtext
\widetext
\begin{equation}
m^2_R(t)= \mu^2_R \left[\frac{T^2_i}{T^2_c}-1\right]
\Theta(-t)
- \mu^2_R \left[1-\frac{T^2_f}{T^2_c}\right] \Theta(t)
\end{equation}
\narrowtext
with $T_i > T_c$ and $T_f \ll T_c$.
With the self-consistent solution and boundary condition for
$t<0$
\widetext
\begin{eqnarray}
[\langle\Phi^2(t<0)\rangle-\langle\Phi^2(0)\rangle]  & = & 0
\; \; ; \; \;
{\cal{U}}^{\pm}_k(t<0)  =  \exp[\mp i \omega_{<}(k)t]
\label{bcmodes}\\
\omega_{<}(k)                & = & \sqrt{\vec{k}^2+m^2_{iR}}
\end{eqnarray}
\narrowtext

This set of Hartree equations is extremely complicated
to be solved exactly.
However it accounts for the process of coarsening\cite{langer1}.
Consider the
equations for $t>0$,
at very early times, when (the renormalized)
$\langle\Phi^2(t)\rangle-
\langle\Phi^2(0)\rangle \approx 0$
the mode functions are the same as in the zeroth order
approximation,
and the unstable modes grow exponentially. By computing the
expression
(\ref{subfi2}) self-consistently  with
these zero-order unstable modes, we see that the fluctuation
operator begins to grow exponentially.

As $(\langle\Phi^2(t)\rangle-\langle\Phi^2(0)\rangle)$ grows
larger,
its contribution to the Hartree equation tends to balance
the negative
mass term, thus weakening the unstabilities, so that only
longer
wavelengths can become
unstable. Even for very weak coupling constants,
the exponentially
growing modes make the Hartree term in the equation of
motion for the mode
functions become large and compensate for the negative mass
term.
Thus when

\[\frac{\lambda_R}{2m^2_{f,R}}\left(\langle\Phi^2(t)\rangle-
\langle\Phi^2(0)\rangle\right)
\approx 1 \]
the instabilities
shut-off, this equality determines the ``spinodal time''
$t_s$.
The modes will still continue to grow further
after this point
because the time derivatives are fairly (exponentially)
large, but eventually
the growth will slow-down when fluctuations sample deep
inside the stable  region.

After the subtraction, and multiplicative renormalization
(absorbed in a
coupling constant renormalization), the composite operator
is finite. The
stable mode functions will make a {\it perturbative}
contribution to the
fluctuation which will be always bounded in time.  The most
important contribution will be that of the {\it unstable
modes}. These will grow
exponentially at early times and their effect will dominate
the dynamics of
growth and formation of correlated domains. The full set of
Hartree equations
is extremely difficult to solve, even numerically, so we
will restrict
ourselves to account {\it only} for the unstable modes. From
the above
discussion it should be clear that these are the only
relevant modes for the
dynamics of  formation and growth  of domains, whereas the
stable modes,  will
always contribute perturbatively for weak coupling after
renormalization.

Introducing the dimensionless ratios (\ref{dimensionless})
in terms of
$m_{f,R}\; ; \; m_{i,R}$, (all momenta are now expressed in
units of
$m_{f,R}$), dividing (\ref{subhartree}) by $m_{f,R}^2$,
using the high temperature
approximation $\coth[\beta_i\omega_{<}(k)/2] \approx
2T_i/\omega_{<}(k)$
for the unstable modes, and expressing the critical
temperature as
$T^2_c=24 \mu^2_R/\lambda_R$, the set of Hartree equations
(\ref{subfi2},
\ref{subhartree}) become the following integro-differential
equation for
the mode functions for $t>0$
\widetext
\begin{equation}
\left[\frac{d^2}{d\tau^2} + q^2 -1 +\int_0^1 dp \left\{
\frac{p^2}{p^2+L^2}\left[{\cal{U}}^+_p(\tau){\cal{U}}^-_p(\tau)-1
\right]\right\}\right]{\cal{U}}^{\pm}_q(\tau)=0 \label{finalhartree}
\end{equation}
\narrowtext
with the boundary conditions (\ref{bcmodes}) for $t<0$ and
\widetext
\begin{equation}
g   = \frac{\sqrt{24\lambda_R}}{4\pi^2}
\frac{T_i}{[T^2_c-T^2_f]^{\frac{1}{2}}}
\label{effectivecoupling}
\end{equation}
\narrowtext

The effective coupling (\ref{effectivecoupling}) reflects
the enhancement of
quantum fluctuations by high temperature effects; for
$T_f/T_c \approx 0$,
and for couplings as weak as $\lambda_R \approx 10^{-12}$,
$g \approx 10^{-7} (T_i/T_c)$. This value of the coupling has
particular significance in inflationary models and arises from bounds
on density fluctuations\cite{brandenberger,kolb,linde}.
The equations (\ref{finalhartree}) may now be integrated
numerically for the
mode functions; once we find these, we can then compute the
contribution of the unstable modes
 to the subtracted
correlation function equivalent to  (\ref{integral})
\widetext
\begin{eqnarray}
{\cal{D}}^{(HF)}(x,\tau)    & = &
\frac{\lambda_R}{6 m_{f,R}^2} \left[\langle \Phi(\vec{r},t)
\Phi(\vec{0},t)\rangle-
\langle\Phi(\vec{r},0)\Phi(\vec{0},0)\rangle\right]
\label{hartreecorr1} \\
3{\cal{D}}^{(HF)}(x,\tau)   & = & g\int_0^1dp
\left(\frac{p^2}{p^2+L^2}
\right)\frac{\sin(px)}{(px)}\left[{\cal{U}}^+_p(t){\cal{U}}^
-_p(t)-1\right]
\label{hartreecorr2}
\end{eqnarray}
\narrowtext
In figure (2) we show $3({\cal{D}}^{HF}(0,\tau)- {\cal{D}}^{HF}(0,0))$
(solid line) and
also for comparison, its zeroth-order counterpart
$3({\cal{D}}^{(0)}(0,\tau)-{\cal{D}}^{(0)}(0,0))$ (dashed
line)
for $\lambda_R = 10^{-12}\; , \; T_i/T_c=2$.
(This value of the initial temperature does not have
any
particular physical significance and was chosen as a
representative).  We clearly see what we expected;
 whereas the zeroth order correlation grows indefinitely,
the Hartree
correlation function is bounded in time and oscillatory. At
$\tau \approx
10.52$ ,  $3({\cal{D}}^{(HF)}(0,\tau)-
{\cal{D}}^{(HF)}(0,\tau))
= 1$,
fluctuations are sampling field configurations near the
classical spinodal, fluctuations
 continue to grow, however,  because the
derivatives are still fairly large. After this
time, the
modes  begin  to probe the stable region in which there is
no
exponential growth. At this point
$\frac{\lambda_R}{2m_{f,R}^2}(\langle\Phi^2(\tau)\rangle-
\Phi^2(0)\rangle)$,
becomes small again because of the small coupling $g \approx
10^{-7}$, and the correction term becomes small.  When
it becomes
smaller than one, the instabilities set in again, the unstable modes
begin to grow and the process repeats.
This gives rise to an oscillatory behavior around
$\frac{\lambda_R}{2m^2_{f,R}}(\langle\Phi^2(\tau)\rangle-
\Phi^2(0)\rangle=1$ as shown in figure (2). We clearly see that
for very weakly coupled theories, the zeroth order correlation function
provides a fairly good approximation to the Hartree correlations up to
the ``spinodal time''. Thus for very weakly coupled theories correlation
functions will be approximately given by (\ref{strucfacxtau},
\ref{strucfactau}) and this permits us to find an approximate result
for the spinodal time (at which fluctuations begin probing the stable
region).

\begin{equation}
\tau_s = \frac{t_s}{\sqrt{2}\xi(0)} \approx -\ln\left[
\left(\frac{3\lambda}{4\pi^3}\right)^
{\frac{1}{2}}\left(\frac{(\frac{T_i}{2 T_c})^3}{[
\frac{T^2_i}{T^2_c}-
1]}\right)\right]
\end{equation}

It is remarkable that the domain size scales as $\xi_D(t)
\approx
t^{\frac{1}{2}}$ just like in classical theories of spinodal
decomposition,
in which the order parameter {\it is not conserved}, as is
the case in this relativistic
scalar field theory,
but certainly for completely different reasons. At the tree
level, we can
identify this scaling behavior as arising from the
relativistic dispersion
relation,  a
situation very different from the classical description of
the Allen- Cahn-
Lifshitz\cite{allen} theory of spinodal decomposition based on a Time-
dependent Landau
Ginzburg model.
 For strong coupling, the Hartree result and the
zeroth-order result depart at very early times\cite{boyshin}.
It is well known within the
context of classical spinodal decomposition that the Hartree
approximation is not correct at intermediate and long times. We are
currently studying a consistent treatment beyond the Hartree
approximation. Details of the calculation and possible extensions will
be presented elsewhere\cite{boyshin}.
Of particular importance will be the study of the interface dynamics,
known to be the relevant description in classical theories\cite{otha}.

\vspace{5mm}

\acknowledgements
I would like to thank D. Jasnow and Y. Oono for very illuminating
discussions and comments. The author has been partially supported by
N.S.F. through grant No: PHY-8921311.

\vspace{5mm}

{\bf Figure Captions:}

\underline{\bf Figure 1:}

$(\tau \exp[-2\tau])\kappa {\cal{S}}(\kappa,\tau)$
for $ \lambda=10^{-12} \; ; \;  T_i/T_c =2 $ and $ \tau =
10,\; 13,\; 16,\; 19$.

\underline{\bf Figure 2:}

 Hartree (solid line) and zero
order (dashed line)
results for
$\frac{\lambda}{2m^2_f}(\langle\Phi^2(\tau)\rangle-
\langle\Phi^2(0)\rangle) = 3 {\cal{D}}(0,\tau)$,
for $\lambda=10^{-12}$, $\frac{T_i}{T_c}=2$.

\end{document}